\documentclass[10pt,preprint2]{aastex}
\usepackage{graphicx}
\usepackage{amsmath} 
\usepackage{natbib}

\newcommand{\beqa}{\begin{eqnarray}}
\newcommand{\eeqa}{\end{eqnarray}}

\def\hMpc{\ifmmode{h^{-1}{\rm Mpc}}\else{$h^{-1}{\rm Mpc}$}\fi}
\def\hkpc{\ifmmode{h^{-1}{\rm kpc}}\else{$h^{-1}{\rm kpc}$}\fi}
\def\hMsun{\ifmmode{h^{-1}M_\odot}\else{$h^{-1}M_\odot$}\fi}

\begin{document}

\title{Feedback in AGN heating of galaxy clusters}

\shorttitle{Feedback in AGN heating of galaxy clusters}
\shortauthors{Hoeft \& Br\"uggen}

\author{M. Hoeft }
\author{M. Br\"uggen } 
\affil{International University Bremen }

\altaffiltext{1}{Campusring 1, 28759 Bremen, Germany }
\altaffiltext{2}{An der Sternwarte 16, 14482  Potsdam, Germany }
\altaffiltext{3}{mhoeft@iu-bremen.de}

\begin{abstract}
One of the challenges that models of AGN heating of the intracluster
medium (ICM) face, is the question how the mechanical luminosity of
the AGN is tuned to the radiative losses of the ICM. Here we implement
a simple 1D model of a feedback mechanism that links the luminosity of
the AGN to the accretion rate. We demonstrate how this simple feedback
mechanism leads to a quasi-steady state for a broad range of
parameters. Moreover, within this feedback model, we investigate the
effect of thermal conduction and find that its relative importance
depends strongly on the cluster mass.
\end{abstract}

\keywords{
galaxies: clusters: general,
galaxies: active,
X-ray: galaxies: clusters	 }

%\twocolumn

\section{Introduction}

\label{sec-intro}

Currently, the most popular model that is invoked to explain the
dearth of gas below about 1 keV in the ICM relies on the heating by a
central AGN (\citealt{binney:95}, \citealt{tabor:93},
\citealt{churazov:01}, \citealt{bruggen:02} , \citealt{bruggen:02b}
\citealt{reynolds:01}). Numerous observations of x-ray deficient
bubbles in clusters and of motions induced by these bubbles have
substantiated this model (e.g. \citealt{fabian:03},
\citealt{mazzotta:02}, \citealt{saxton:01}, \citealt{mcnamara:01},
\citealt{blanton:01}). On the theoretical side, the efficiencies with
which bubbles heat the ICM has been investigated, both numerically and
analytically. E.g. \citet{ruszkowski:04} found that
up to 50 \% of the internal energy of these bubbles is dissipated
through viscous dissipation of waves and $pdV$ work. The heating
profile caused by bubbles has been parametrized in a prescription
named `effervescent heating' by \cite{begelman:01}, which will be used
later in this paper.

However, one of the challenges that models of AGN heating face, is the
question how the mechanical luminosity of the AGN is fine-tuned to the
radiative losses of the ICM. If AGN heating is responsible for a
diminution of the mass deposition rate in a whole range of clusters,
the AGN needs to be regulated by the ICM itself. It has been suggested
by some authors that AGN feedback may play a crucial role in
self-regulating cooling flows (e.g. \citealt{churazov:02},
\citealt{ruszkowski:02}, \citealt{brighenti:03}).

In this paper we present a simple model by which the central AGN
adjusts its luminosity. This model assumes that the mechanical
luminosity of the AGN is proportional to the accretion rate onto the
AGN. In this picture, a very luminous central source will lead to a
decrease in the accretion rate and hence also to a decline of the
luminosity until, eventually, the cluster reaches a quasi-steady
state. With the aid of a spherically symmetric 1D model we study
under which circumstances a self-regulated cooling flow is
established. For feedback to be a viable model, the parameters of this
model should not need any tuning.

In the following we will describe our initial models, our
computational method and the implementation of feedback in more
detail. Then, in section 4, we will present and discuss our results.

%%%%%%%%%%%%%%%%%%%%%%%%%%%%%%%%%%%%%%%%%%%%%%%%%%%%%%%%%

\section{Model}

\label{sec-ini}

\subsection{Initial profiles}

The initial cluster profiles are computed within the present canonical
cosmological parameters, see Tab.~\ref{tab-parameter}. For the dark matter
density distribution we assume an NFW profile given by
(\citealt{navarro:96})
\begin{equation}
  \rho_{\rm dm}
  =
  \delta_c \: 
  r^{-1} \: 
  ( r_s + r )^{-2}
  ,
\end{equation}	
where $\delta_c$ is the characteristical density parameter 
and $r_s$ the scaling radius. Thus, a cluster model is
characterized by the virial mass $M_{\rm vir}$ and by the
concentration parameter $c=r_s/r_{\rm vir}$. Note, that the virial
radius can be derived from $4/3 \; \pi r_{\rm vir}^3 \, h^2 \,
\rho_{\rm crit} \Delta_c = M_{\rm vir}$, where $\rho_{\rm crit}$ is
the critical density and the density contrast $\Delta_c$ can be
approximated by $\Delta_c = 178\, \Omega_0^{0.45}$ \citep{eke:98}.

Unlike for dark matter distributions, there is no universal profile
for the intracluster gas. Here, we follow \citet{roychowdhury:04} and
assume that the initial temperature distribution is given by
(\cite{loken:02})
\begin{equation}
  T_{\rm initial}(r) 
  =
  1.3 \:
  T_{\rm ew}\:
  \left( 1 + 1.5 \, r/r_{\rm vir} \right)^{-1.6}
  .
  \label{eq-init-temp}
\end{equation}  
Now, the emission-weighted X-ray temperature, $T_{\rm ew}$, is known
to scale with the cluster mass. Here, we adopt the $T_X-M$-relation by
\cite{sanderson:03} who have analysed a large sample of virialized
systems:
\begin{equation}
  M_{200}
  =
  2.34\times10^{13} M_{\odot} \:
  \left( \frac{ k_B T_{\rm ew } }{ 1 \: {\rm keV} }  \right)^{1.84}
  .
\end{equation}
Moreover, we assume that the gas is in hydrostatic equilibrium, i.e. the
pressure force is balanced by gravitation
\begin{equation}
  \frac{1}{\rho_{\rm gas}}
  \frac{{\rm d} P}{ {\rm d} r }
  =
  -
  \frac{G M_{\rm total} (<r)}{ r^2 }
  \label{eq-pressure}
  ,
\end{equation}
where $G$ is the gravitational constant and $M_{\rm total}(<r)$ is the
total cumulative mass up to $r$. We can express the pressure as $P=n
k_B T$, where $n$ is the particle number density, which is given for a
fully ionized gas of primordial composition by $n=n_{\rm HII} + n_{\rm
HeIII} + n_e = n_{\rm H} (8-5Y_{\rm He})/(4-4Y_{\rm He})$, where
$Y_{\rm He}$ is the Helium mass fraction. Since we assume an initial
temperature profile, we can rewrite Eq.~(\ref{eq-pressure}) and solve
for the Hydrogen density $n_{\rm H}$
\begin{equation}
  \frac{ {\rm d} n_{\rm H} }{ {\rm d} r }
  =
  -
  \frac{ n_{\rm H} }{ T }
  \left\{
	 \frac{\mu m_{\rm H} } { k_B }
	 \frac{1}{r^2}
	 M(<r)
	 +
    \frac{ {\rm d} T}{ {\rm d} r }
  \right\}
  \label{eq-initial-inte}
  ,
\end{equation} 
where the mean molecular weight is given by $\mu = 4 / ( 8-5Y_{\rm
He})$. We integrate Eq.~(\ref{eq-initial-inte}) subject to the
condition that at radius $r_{200}$ the ratio of the cumulative masses
$M_{\rm gas}/M_{\rm dm}$ is equal to the cosmological value
$\Omega_B/(\Omega_0 - \Omega_B)$.

%%%%%%%%%%%%%%%%%%%%%%%%%%%%%%%%%%%%%%%%%%%%%%%%%%%%%%%%%

\subsection{Time integration}

The entropy index $\sigma = T / n_e^{2/3}$ at a given radius evolves
with time according to
\begin{equation}
  \frac{ {\rm d} \sigma(r) }{ {\rm d} t }
  =
  \sigma(r) \:
  \frac{2}{3}
  \frac{1}{P(r)}
  \{ {\cal H}(r) - \Gamma(r) \}
  ,
\end{equation}
where ${\cal H}(r)$ denotes the total heating and $\Gamma(r)$ the
cooling cooling rate. For the subsequent calculations it is convenient
to express the entropy profile as a function of the total enclosed
mass, i.e. $\sigma=\sigma(M)$. Starting from an initial entropy
profile that is calculated from the models described above, the
cluster is evolved in time.

After each time step, $\Delta t$, the new profiles are calculated in two
steps: first we calculate the change of entropy subject to heating and
cooling, i.e.
\begin{equation}
	\sigma 
	\rightarrow 
	\sigma 
	+ 
	\frac{ {\rm d} \sigma(r) }{ {\rm d} t }\Delta t 
	\ .
\end{equation}
In the second step we calculate the new hydrostatic
equilibrium into which the halo settles by solving
\begin{eqnarray}
  \frac{ {\rm d} r }{ {\rm d} M }
  &=&
  \frac{1}{4 \pi r^2\rho_{\rm gas}(r)} =  \frac{1}{4 \pi r^2}\left (\frac{\sigma}{P}\right )^{1/\gamma}
  \nonumber
  \\
  \frac{{\rm d} P }{ {\rm d } M }
  &=&
  -
  \frac{ G M_{\rm total}(<r) } { 4 \pi r^4 }
  \label{eq-radius-pressure-inte}
  .
\end{eqnarray}
Equations~(\ref{eq-radius-pressure-inte}) are discretized on a non-uniform
grid. A fine grid covers the inner region of the cluster out to a mass
that in the initial model corresponds to a radius of $5\times 10^{-2} 
r_{\rm vir}$ and a coarser grid that covers the outer region out to
$r_{200}$. Thus, the mass of a shell in the finely resolved region is
0.1\% of the cumulative mass $M_{\rm gas}(<r)$ compared to 2\% in the
coarse region. This results in about 4000 mass shells altogether.
 
Equation~(\ref{eq-radius-pressure-inte}) is solved subject to the
boundary condition that the pressure at $r_{200}$ remains constant during
the entire evolution, i.e. $P(r_{200})=P_{200}$. In practice, we solve
the system of coupled differential equations
(\ref{eq-radius-pressure-inte}) subject to the initial conditions:
\begin{eqnarray}
  M(0)
  &=& 0
  \nonumber
  \\
  P(0)
  &=&
  P_0
  \label{eq-init-cond}
  ,
\end{eqnarray}
where $P_0$ is a trial value which is varied until the pressure at
$r_{200}$ is equal to the specified pressure $P_{200}$.  Note that
this implies that the radius $r_{200}$ changes in the course of time
as the cluster expands and contracts.

%%%%%%%%%%%%%%%%%%%%%%%%%%%%%%%%%%%%%%%%%%%%%%%%%%%%%%%%%

\subsection{Radiative cooling and conduction}

\label{sec-lambda-conduct}

In the temperature range that we are mainly interested in, namely
between $\rm 10^5 - 10^8\:K$, radiative cooling is dominated by
free-free emission. For a fully ionized plasma the cooling rate can be
approximated by \citep{katz:96}
\begin{equation}
  \Lambda_{\rm free-free}
  =
  1.42 \times 10^{-27} \: 
  g_{\rm ff} \:
  T^{1/2}
  ( n_{\rm H} + 4n_{\rm He} ) n_e \;
  \frac{ \rm erg}{ \rm s \, cm^3 } 
  ,
\end{equation}  
with the Gaunt factor
\begin{equation}
  g_{\rm ff}
  =
  1.1 + 0.34 \exp \{ - (5.5 - \log T )^2 / 3 \}
  .
\end{equation}
In addition we consider metal-dependent cooling,
which we approximate crudely following \citet{theis:92}
\begin{eqnarray}
  \Lambda_{\rm metal} 
  &=&
  10^{-22-5m(Z)+7\sqrt{Z}} \:
  T^{m(Z)} \:
  n_{\rm H}^2 \;
  \frac{ \rm erg}{ \rm s \, cm^3 } 
  \nonumber
  \\
  m(Z) 
  &=&
  \frac{2.5+7\sqrt{Z}}{5-\log(1.48\times 10^{11} Z^{1.1} + 10^6)}  
  ,
\end{eqnarray}
where $Z$ denotes the mass fraction of metals. Here we assume
$Z=0.3\:Z_\odot$ and we limit the cooling in the core by assuming a 
 minimum entropy index 
$\sigma_{\rm min} = 10^{-2}\:{\rm keV \, cm^2}$.

The role of thermal conduction in the ICM has been the subject of a
long debate and, owing to the complex physics of MHD turbulence, the
value of the effective conductivity remains uncertain. The thermal
conductivity of an unmagnetised, fully ionised plasma was calculated
by \citet{spitzer:62}. Originally it was thought that the magnetic
field in clusters strongly supressed the thermal conductivity because
the magnetic fields prevented an efficient transport perpendicular to
the field lines. Even if the transport can be efficient along the
magnetic field lines, the overall isotropic conductivity was thought
to be many orders of magnitude less than the Spitzer value. This
paradigm has been supported by a number of observations, such as sharp
edges at so-called cold fronts and small-scale temperature variations
(\citealt{markevitch:00}, \citealt{vikhlinin:01}).

Recent theoretical work by \cite{narayan:01}, \cite{chandran:99},
\cite{chandran:98} and earlier work by \cite{rechester:78} have shown
that a turbulent magnetic field is not as efficient in suppressing
thermal conduction as previously thought. It is argued that chaotic
transverse motions of the tangled magnetic field lines can enhance the
cross-field diffusion to an extent that the effective conductivity is
of the order of the Spitzer value. Thus, we decided to investigate the
effects of thermal conduction within our feedback model. Recently, the
role of thermal conduction in clusters was also investigated
numerically by \cite{dolag:04} and \cite{jubelgas:04}.

The energy flux due to thermal conduction is given by
\begin{equation}
  F_{\rm cond}
  =
  - \kappa \:
  \nabla T
  ,
\end{equation}  
where $\kappa$ is the coefficient of thermal conductivity which we
assume to be a fraction $f$ of the Spitzer conductivity
\begin{equation}
  \kappa_0 
  T^{5/2}
  \approx
  5 \times 10^{-7} \:
  T^{5/2} \;
  \frac{ \rm erg }{ \rm s \, cm \, K } \;
  .
\end{equation}
Under the assumption of spherical symmetry, the heating and cooling
rate due to thermal conduction become
\begin{eqnarray}
  \Gamma_{\rm cond}
  &=&
  -
  \frac{1}{r^2}
  \frac{ {\rm d} }{ {\rm d} r }
  \left\{
    r^2
    F_{\rm cond}
  \right\}
  \\
  &=&
  - f_{\rm Spitz} \kappa_0
  \frac{ T^{3/2} } { r }
  \left\{
	 2 T \frac{ {\rm d} T }{ {\rm d} r }  
	 +
	 \frac{5}{2} r \left( \frac{ {\rm d} T }{ {\rm d} r }	\right)^2
	 +
	 r T \frac{ {\rm d}^2 T }{ {\rm d} r^2 }
  \right\}
  \nonumber
  .
\end{eqnarray}  
The timestepping in the integration of
Eq. (\ref{eq-radius-pressure-inte}) is constrained by the appropriate
Courant condition \citep{ruszkowski:02}. The most restrictive term
comes from thermal conduction which can be obtained from a von Neumann
stability analysis, i.e.
\begin{equation}
\Delta t
  =
  \frac{ 3 k_B n (\Delta r)^2 }
       {4 f_{\rm Spitz} \kappa_0 T^{5/2} }
 ,
\end{equation}
where $\Delta r$ denotes the local spatial resolution. The time step
obviously changes in the course of the simulation but for our
resolution a typical time step is of the order of $0.05\:{\rm Myr}$.

Since we do not use a uniform grid, especially with respect to the 
spatial spacing, we compute the derivatives of the temperature by 
a local polynomial fit.  Around each
point in radius we expand the temperature to second order
\begin{equation}
  T(r)
  =
  T(r_0)
  +
  \frac{ {\rm d} T }{ {\rm d} r } ( r - r_0 )
  +
  \frac{1}{2}  
  \frac{ {\rm d}^2 T }{ {\rm d} r^2 } ( r - r_0 )^2
\end{equation}  
and approximate the derivatives with a weighted least-square
regression. We solve the equation
\begin{equation}
  {\bf Q } 
   =
  { \bf W } 
  \cdot 
  \left(
    \begin{array}{c}
	   T(r_0)  \\
	   {\rm d} T / {\rm d} r  \\
	   1/2 \: {\rm d}^2  T / {\rm d} r^2  
	 \end{array}
  \right)	 
\end{equation}
where
\begin{eqnarray}
  Q_i
  &=&
  \Sigma_n T_n \omega_{n} (r_n-r_0)^{i}
  \nonumber
  \\
  W_{ij}
  &=&
  \Sigma_n \omega_{n} (r_n-r_0)^{i+j}
  .
\end{eqnarray}
The sum is taken over all sampling points $n$ and $i,j \in \{ 0, 1, 2 \}$. The weights are given by
\begin{equation}
  \omega_{n} 
  =
  \left\{
	 \begin{array}{r@{\quad:\quad}l}
	   (r_n - r_{n-1})\:e^{ -  4 ( r_n - r_0 )^2 / \Delta^2 }  &  |r_n-r_0| < 2 \Delta  \\
		 0  & {\rm otherwise } 
	 \end{array}
   \right.
\end{equation}  		
For the smoothing length we take $\Delta = \max \{ 0.05 \times r, 0.05 \:
{\rm kpc } \}$. Finally, we neglected convection which proved to be virtually
irrelevant for our simulations since the entropy remained monotonous
even in the presence of strong heating.

%%%%%%%%%%%%%%%%%%%%%%%%%%%%%%%%%%%%%%%%%%%%%%%%%%%%%%%%%

\subsection{Heating and feedback}

In the centres of clusters active galactic nuclei inflate buoyant bubbles of
relativistic gas which release some of their internal energy by
$pdV$-work on the ambient medium. This work is done as the bubbles
expand on their ascent through the stratified cluster
medium. \cite{churazov:01} and \cite{begelman:01} have calculated this
work and found that the heat released by one bubble is given by
\begin{equation}
h(r) 
	\propto 
	P^{(\gamma_b - 1)/\gamma_b} \;
  \frac{1}{r} \;
  \frac{ {\rm d } \, \ln P } { {\rm d} \, \ln r }  
.
\end{equation}
Under the assumption of spherical symmetry the heat has to be
distributed over spherical shells, which introduces a geometric factor
$f_{\rm geom} \propto 1\;/\;4\pi r^2$. However, at radii smaller than
a typical bubble diameter this argument does not hold. In the center
we assume that the heating is given by the heating function of one
bubble, therefore we cut off the geometrical factor below the typical
bubble size $d_{\rm bubble}$, by changing $f_{\rm geom}$: $r^2 \to r^2
+ d_{\rm bubble}^2$. As bubbles rise through the ICM they are likely
to be disrupted by Rayleigh-Taylor instabilities. The smaller
fragments rise more slowly and are disrupted more easily until they
virtually stop. As a result the effervescent heating rate is
effectively cut off at the periphery of the cluster. This effect is
approximated by introducing a factor $\exp\{ - (r/r_{\rm disr}) \}$,
where the disruption radius, $r_{\rm disr}$, is set to 0.5 Mpc. We
note, however, that this cut-off at large radii only has a mild effect
on our simulations and is not strictly necessary. Thus the
effervescent heating rate is given by
\begin{equation}
  {\cal H }(t) 
  = {\cal L}_{\rm eff}(t) 
  h(r) 
  f_{\rm geom}(r), 
\end{equation}
where the radially dependent functions are normalized according to
\begin{equation}
  \int\nolimits_{r_{\rm min}}^{r_{\rm max}} 
  {\rm d} r \; 4\pi r^2 \; 
  h(r) \;  
  f_{\rm geom}(r) 
  = 
  1 \ .
\end{equation}
The geometrical factor is given by
\begin{equation}
  f_{\rm geom}(r) 
  =  
  \frac{ \exp \left\{ -( r/r_{\rm disr}) \right\}}{ 4 \pi ( r^2  + d_{\rm bubble}^2 ) }
  ,
\end{equation} 
and ${\cal L}_{\rm eff}(t)$ denotes the total `effervescent' luminosity at
time $t$. 

In our continuous feedback model we assume that the luminosity of the
AGN is proportional to the mass accretion onto the core, i.e.
\begin{equation}
  {\cal L}_{\rm eff}(t)
  = 
  \epsilon  \dot{M}_{\rm core}c^2
  ,
\end{equation}
where the core is made up of gas whose entropy index lies below a
minimal value, $\sigma_{\rm min}$.

The accretion rate is calculated by a linearly weighted least-square
regression of the function $M_{\rm core}(t)$, where a similar
procedure as decribed in Sec.~\ref{sec-lambda-conduct} is used with a
smoothing timescale of 20~Myr. In order to improve numerical stability
we introduce a delay time of 2~Myr, i.e. the luminosity is
proportional to the accretion rate of 2~Myr ago.

\section{Results}

\subsection{Self-regulated luminosities}

As our canonical model we take a cluster with a dark matter mass of
$M_{\rm vir} = 4\times10^{14}\:M_\odot$ and a concentration parameter
$c=4$. These values are chosen to resemble those of the cooling flow
cluster Abell~2052 (\citealt{blanton:03}). We start all our
simulations from the initial temperature profile given by
Eq.~(\ref{eq-init-temp}) and leave conduction and heating switched off
until the cluster has formed a cold core. Only then we begin with the
simulation of feedback.\\

As the ICM cools and the central cooling times decrease, the accretion
onto the core increases and so does, according to our prescription for
feedback, the luminosity of the AGN. We will start by first describing
the simulation without thermal conduction.

Fig.~\ref{fig-L_efferv_from_feedback} shows the self-regulated
luminosity of the effervescent heating as a function of time for
various values of the efficiency, $\epsilon$. After an initial phase
which only lasts for a few hundred Myrs the luminosity converges to a
value of around $L\sim10^{44}$ erg s$^{-1}$ and remains steady for the
entire length of the simulation, which is longer than 3~Gyrs.  Runs
with different efficiencies show, that the effervescent heating
is in fact the regulating parameter. Even for quite different
efficiencies that span two orders of magnitude the luminosities end up
at quite similar values.

\begin{figure}
\begin{center}
\includegraphics[width=0.5\textwidth,angle=-90]{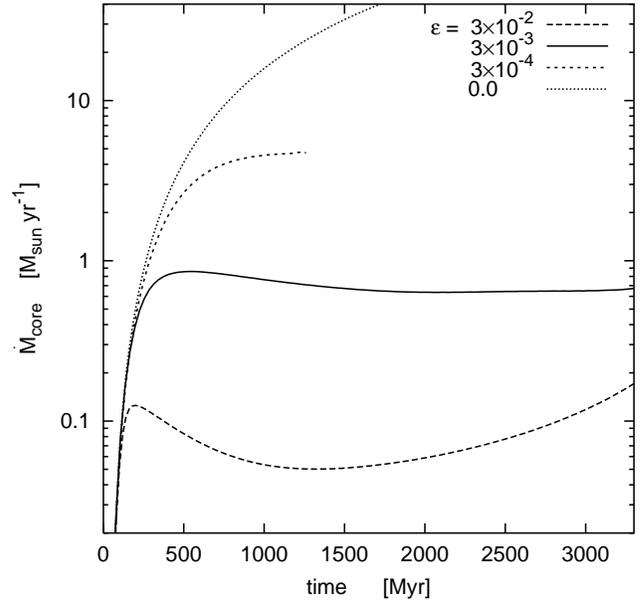}
\caption{Mass accretion rate onto the core as a function of time.  The
lines correspond to models with different efficiencies,
$\epsilon$. The oscillations that appear at late times for very low
luminosities are numerical artefacts caused by the finite resolution
of the mass shells. The thick double-dashed line that is labelled by
``all off'' is the accretion rate in a model without heating. It shows
how quickly the mass accretion rate diverges in the absence of
heating.}\label{fig-accretion_from_feedback}
\end{center}
\end{figure}

The heating stalls the accretion onto the core as is shown in
Fig.~\ref{fig-accretion_from_feedback} which shows the accretion rate
as a function of time. One can see how the accretion rate adjusts
itself to a value just below 1 $M_{\odot}$/yr after a few hundred
Myrs. For comparison we also show the corresponding scenario with
neither heating nor conduction by the thick double-dashed line that is
labelled by ``all off''. It shows how quickly the accretion rate
diverges in the absence of heating.

Many AGN are believed to be recurrent with duty cycles between 10 and
a few 100 Myrs. This suggests that a non-linear feedback process is at
work, i.e. that the luminosity is not linearly related to the
accretion rate as presumed here. However, nothing is known about the
parameters of such a non-linear feedback mechanism. One could, for
instance, introduce upper and lower thresholds for $\dot{M}$, such
that the AGN switches on if the accretion rate lies above the upper
threshold and is switched off when it falls below the lower
threshold. Depending on the values for these thresholds one can
reproduce sensible duty cycles, but since one has introduced two more
parameters, this is no great achievement. Therefore, we decided to
focus on a simple prescription for continuous feedback which
introduces only a single parameter. Most importantly, our results
depend only very weakly on this parameter. Furthermore, even if in nature
feedback operates in some non-linear fashion, our results are still
valid in a time-averaged (i.e. averaged over a few duty cycles) sense
since the recurrence times are small compared to the time scales
considered here. Thus, if the AGN is active for only part of the time,
the mechanical luminosities of the central AGN can be higher than the
$\sim 10^{44}$ erg s$^{-1}$ found here.

\subsection{The effect of conduction}

\begin{figure}
\begin{center}
\includegraphics[width=0.5\textwidth,angle=-90]{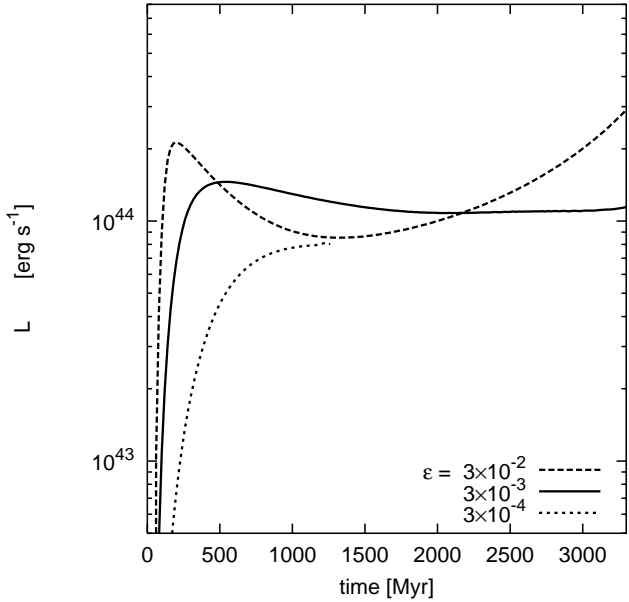}
\caption{Luminosity of the effervescent heating as a function of time.
The lines correspond to models with different efficiencies,
$\epsilon$, The oscillations that appear at late times for very low
luminosities are numerical artefacts caused by the finite resolution
of the mass shells.}
\label{fig-L_efferv_from_feedback}
\end{center}
\end{figure}

\begin{figure}
\begin{center}
\includegraphics[width=0.5\textwidth,angle=-90]{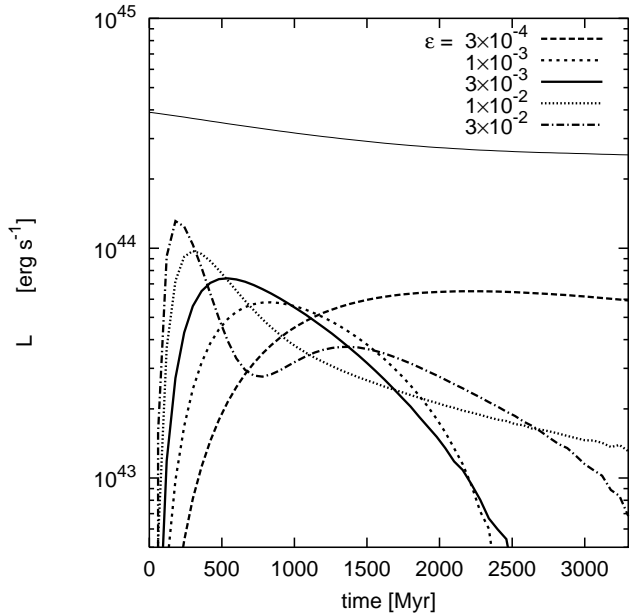}
\caption{Same as Fig.~\ref{fig-L_efferv_from_feedback} but with
thermal conduction. For comparison, the thin line denotes the volume
integrated radiative luminosity of the ICM as a function of time.}
\label{fig-L_cond}
\end{center}
\end{figure}

Next, we studied the effects of thermal conduction on our models.
Fig.~\ref{fig-L_cond} is the corresponding plot to
Fig.~\ref{fig-L_efferv_from_feedback} with thermal conduction (where
we assumed $f=0.3$). The solid lines in
Figs.~\ref{fig-L_efferv_from_feedback} and \ref{fig-L_cond} both
correspond to the same efficiency of $3\times 10^{-3}$ and it is
evident that the presence of thermal conduction leads to decrease of
the accretion rates and, consequently, also of the effervescent
luminosities. Initially, i.e. after about 500 Myrs, the luminosities
are very similar at $\sim 10^{44}$ erg s$^{-1}$ but at later times
thermal conduction suppresses the luminosities to less than $\sim
10^{43}$ erg s$^{-1}$. Higher efficiencies reduce the power of the AGN
even further, while efficiencies of around $3\times 10^{-4}$ result in
a very steady luminosity. In any case, for a wide range of
efficiencies we obtain cluster models that remain stable over long
times and require 'effervescent' luminosities of not more than $\sim
10^{44}$ erg s$^{-1}$.\\

\begin{figure*}
\begin{center}
\includegraphics[width=0.5\textwidth,angle=-90]{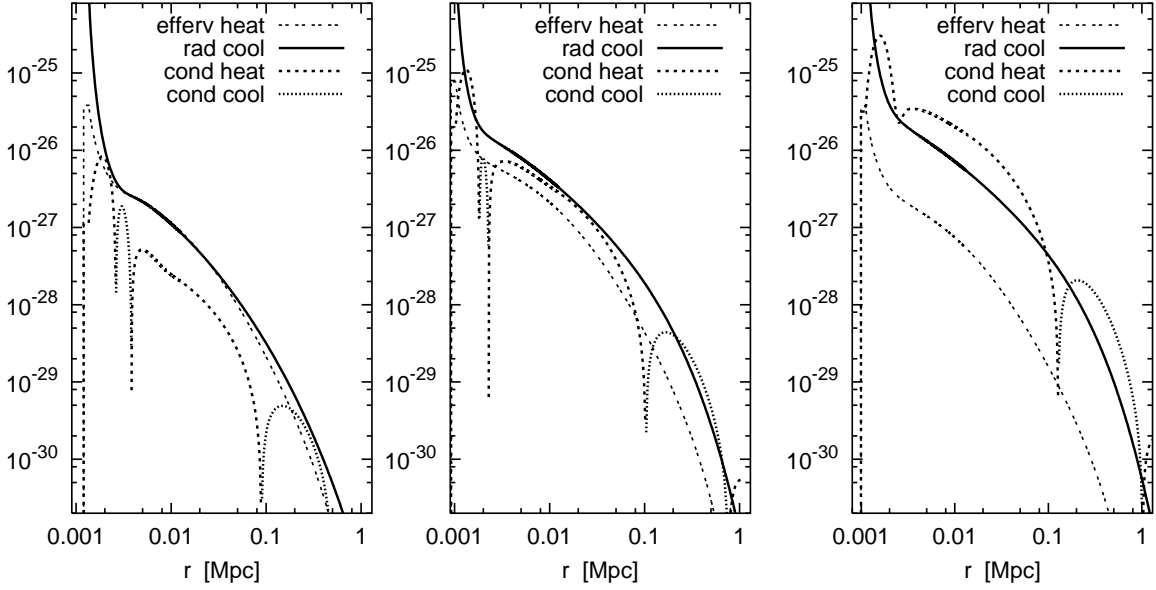}
\caption{The volume heating and cooling rates for three models of
different masses (from left to right the mass is $10^{14}\:M_\odot$,
$4\times10^{14}\:M_\odot$ and $10^{15}\:M_\odot$; $c=4$ for all
models). The units on the vertical axis are erg cm$^{-3}$s$^{-1}$.
All three figures correspond to a time of 1~Gyr after the AGN has been
switched on. The feedback parameter is $\epsilon = 3\times 10^{-3}$.}
\label{fig-heat-cool}
\end{center}
\end{figure*}

Fig.~\ref{fig-heat-cool} shows the volume heating and cooling rates of
our simulations of clusters with masses of $10^{14}\:M_\odot$,
$4\times10^{14}\:M_\odot$ and $10^{15}\:M_\odot$ with an efficiency
parameter of $\epsilon = 3\times 10^{-3}$ and an assumed conductivity
of 30\% of the Spitzer conductivity ($f=0.3$). The figure shows the
curves at a time of 1~Gyr after the start of the feedback. The solid
line shows the radial cooling profile, while the double-dashed line
shows the effervescent heating rate. It is evident from the left panel
that for the cluster with a mass of $10^{14}\:M_\odot$ the radiative
cooling is almost exactly balanced by the effervescent heating while
conduction has very little effect on the energy budget. Also note,
that thermal conduction can, both, heat and cool the cluster as is
shown by different lines in Fig.~\ref{fig-heat-cool}. In the outer
regions and at the edge of the core, conduction leads to cooling while
regions between a few kpc to $\sim 100$ kpc are heated. This has also
been inferred from observations of M87 \citep{ghizzari:04}.
In the very center the heating due to conduction
drops to small values since the temperature is very low. However, the
relative importance of conduction increases with mass. This is not
unexpected since the thermal conductivity depends strongly on
temperature and temperature scales with mass. For the cluster with
$M=4\times10^{14}\:M_\odot$ conduction is already more important than
effervescent heating for most regions within the cluster. Especially
conduction can heat the region around the core and thus effectively
suppresses the accretion. The figure shows that conduction alone
almost compensates the radiative cooling, particularly in regions
where the bulk of X-ray emission comes from. This is in agreement with
previous findings e.g. by \citet{narayan:01} and \citet{fabian:02}.

\subsection{Comparisons with observations}

While we do not intend to undertake a thorough modelling of the observed
features of cooling flow clusters, it is still essential to verify
whether our model can reproduce the gross features of typical cooling
flow clusters. 

\begin{figure}
\begin{center}
\includegraphics[width=0.5\textwidth,angle=-90]{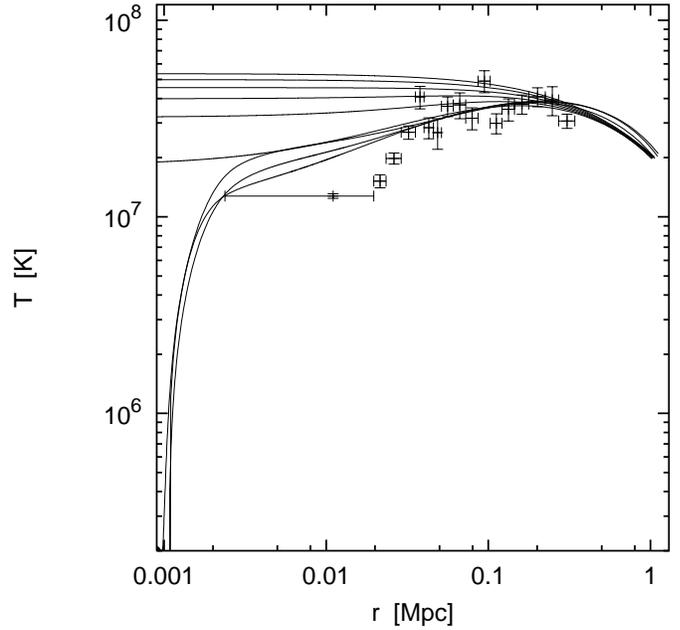}
\caption{Temperature profiles for a cluster model with $M_{\rm vir} =
4\times10^{14}\:M_\odot$ and $c=4.0$. Mass and concentration are
chosen close to the values of Abell~2052, which is a well known
cooling flow cluster. The different lines correspond to different
points in time with a new line drawn every 2~Gyr up to 16~Gyr. Symbols
indicate the observed temperature profile \citep{blanton:03}.  }
\label{fig-temp_profiles}
\end{center}
\end{figure}

Fig.~\ref{fig-temp_profiles} shows the temperature profiles for a
model with $M_{\rm vir} = 4\times10^{14}\:M_\odot$ and $c=4.0$ at
different times.  Mass and concentration parameter are chosen to lie
close to the values of Abell~2052, which is a well known cooling flow
cluster. The initial temperature profile is fairly flat as shown by
the top curve in Fig.~\ref{fig-temp_profiles}, but slowly the central
temperature drops and after about 9 Gyrs a cold core has formed. The
symbols in Fig.~\ref{fig-temp_profiles} denote the observed
temperatures for A~2052 \citep{blanton:03}. We note that the data are
consistent with our model at late times, i.e. after the formation of
the core. Especially the slopes within 50 kpc agree nicely, whereas in
the outer regions the observations show higher temperatures than our
model. This may be an indication that in the periphery other heating
processes are at work. Given the simplicity of our model, it may be
imprudent to overstretch its similarity with observations but in any
case it is encouraging that the profiles that we obtain are not
disparate from observations. Again, we wish to stress that the
temperature profiles are quite insensitive to the efficiency of the
feedback.

\begin{figure}
\begin{center}
\includegraphics[width=0.5\textwidth,angle=-90]{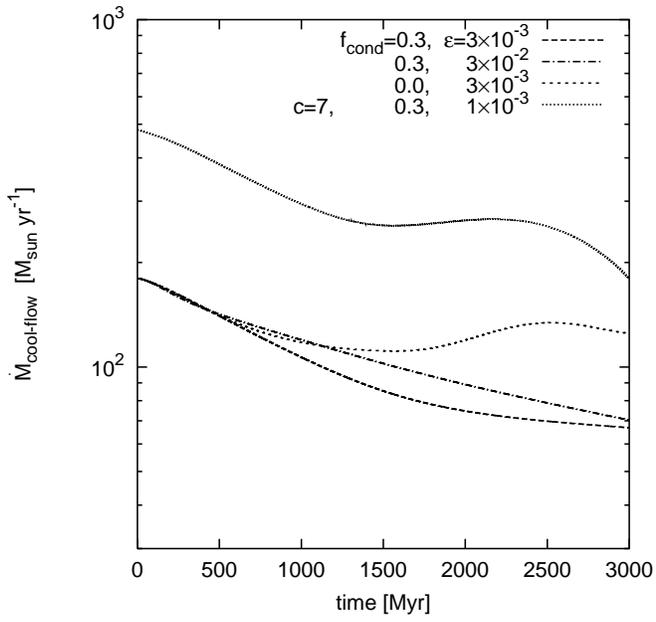}
\caption{Mass deposition rates inferred according to the classical
cooling flow model, i.e. Eq.(\ref{eq-cool-flow}).  The dashed lines
correspond to our canonical model ($M_{\rm vir} =
4\times10^{14}\:M_\odot$ and $c=4$) with different efficiencies and
conductivities, while the dotted line corresponds to a model with
$\epsilon=10^{-3}$ but a higher concentration parameter
($c=7$).}\label{fig-class_cool_flow}
\end{center}
\end{figure}

Finally, it is interesting to compare our accretion rates to the
accretion rates inferred from the total x-ray luminosity according to
the classical cooling flow model, given by
\begin{equation}
\dot{M}_{\rm cool-flow} 
	\approx
	\frac{2}{5}
	\frac{\mu m_{\rm H}}{k_B T_X}
	L_X (<r_c),\
\label{eq-cool-flow}
\end{equation}
where $r_c$ is the cooling radius at which $t_c \approx H_0^{-1}$ and
$L_X$ is the bolometric x-ray
luminosity. Fig. \ref{fig-class_cool_flow} shows the inferred mass
deposition rates for some of our simulated clusters. Comparing
$\dot{M}_{\rm cool-flow}$ to $\dot{M}_{\rm core}$ reveals that the
accretion rates inferred from the X-ray emission are considerably
higher than the rates derived from our single-phase model. In the
presence of heating, the simple equation for $\dot{M}_{\rm cool-flow}$
overestimates the accretion onto the core at least by a factor of
10. The reason is, of course, that the radiative losses are
replenished by bubble heating and thermal conduction, and thus
unaccounted for in Eq. (\ref{eq-cool-flow}).  Again, the variation of
$\dot{M}_{\rm cool-flow}$ with $\epsilon$ is relatively weak. To put
this variation into perspective we show how much more strongly
$\dot{M}_{\rm cool-flow}$ depends on the concentration parameter (top
curve in Fig. \ref{fig-class_cool_flow}) when all other parameters
remain the same.

\subsection{Summary}

We found that a simple prescription for feedback leads to a
self-regulated steady-state model for the ICM: For a wide range of
efficiencies the self-regulated luminosities lie around $L\sim
10^{44}$ erg/s which corresponds to a mass accretion rate of
$\dot{M}\sim 1 M_{\odot}$/yr. Thermal conduction decreases the
luminosity in a self-regulated model and is the more important the
more massive the cluster is.

The resulting profiles for temperature are consistent with
observations of cooling flow clusters. In the presence of heating, the
classical cooling flow model grossly overestimates the mass deposition
rates.\\

\noindent {\sc acknowledgement }

The authors wish to thank Mateusz Ruszkowski and
Eugene Churazov for many helpful suggestions and Sebastian Heinz for
his early involvement in the project. Support by DFG through grant BR
2026/2 is gratefully acknowledged. The calculations were performed at
the {\sc Clamv} at the International University Bremen.

\bibliography{cooling-flow}
\bibliographystyle{apj}

\begin{table}
\begin{center}
\begin{tabular}{clcc}
\hline
\multicolumn{2}{l}{Cosmological model} \\
 & matter density & $\Omega_0$   &  0.3 \\
 & baryon density & $\Omega_B$   &   0.04 \\
 & Hubble constant    &   $h = H_0 / H_0^{100}$         &  0.7  \\
 & Helium mass fraction & $Y_{\rm He}$ & 0.24 \\[.6ex]
\multicolumn{2}{l}{Dark matter halo} \\ 
 & dark matter virial mass $^\ast$ & $M_{\rm vir}$ &  $10^{14}\:M_\odot$ \\
 & concentration parameter $^\ast$  &  c            &  4  \\[.6ex]
\multicolumn{2}{l}{Effervescent energy feedback}  \\
 & efficiency  $^\ast$       & $\epsilon$         & $3\times 10^{-3}$  \\
 & initial bubble size      & $d_{\rm bubble}$   & $20\:{\rm kpc}$  \\
 & disruption radius        & $r_{\rm disr}$     & $500\:{\rm kpc}$  \\[.6ex]
\multicolumn{2}{l}{Conduction} \\
 & Spitzer fraction $^\ast$ & $f_{\rm Spitz}$    & 0.3  \\  	
\hline 
\end{tabular}
\end{center}
\caption{
	Parameters used for our standard simulation. To study other
cluster models parameters marked with $^\ast$ are varied.
}
\label{tab-parameter}
\end{table}

\end{document}